\documentclass[11pt]{article}

\usepackage{amsmath, amssymb}
\numberwithin{equation}{section}

\usepackage{graphicx}

\usepackage{color}

\newcommand{\p}{\partial}
\newcommand{\e}{\varepsilon}
\newcommand{\lm}{\Lambda}

\newcommand{\N}{\mathbb{N}}
\newcommand{\Z}{\mathbb{Z}}
\newcommand{\R}{\mathbb{R}}
\newcommand{\C}{\mathbb{C}}

\newcommand{\A}{\mathcal{A}}

\newcommand{\D}{\mathcal{D}}

\newcommand{\M}{\overline{\mathcal{M}}}
\renewcommand{\H}{\mathcal{H}}
\renewcommand{\P}{\mathcal{C}}
\renewcommand{\d}{\mathrm{d}}

\newtheorem{dfn}{Definition}[section]
\newtheorem{lem}[dfn]{Lemma}

\newtheorem{thm}[dfn]{Theorem}
\newtheorem{rmk}[dfn]{Remark}

\newtheorem{emp}[dfn]{Example}
\newtheorem{cnj}[dfn]{Conjecture}

\newenvironment{prf}{\noindent {\it Proof} \ }{\hfill $\Box$}

\DeclareMathOperator{\Res}{\mathrm{Res}}
\DeclareMathOperator{\ad}{\mathrm{ad}}

\begin{document}

\title{Fractional Volterra Hierarchy}
\author{Si-Qi Liu \quad Youjin Zhang \quad Chunhui Zhou \\
{\small Department of Mathematical Sciences, Tsinghua University}\\
{\small Beijing 100084, P. R. China}}
\date{}\maketitle

\begin{abstract}
The generating function of cubic Hodge integrals satisfying the local Calabi-Yau condition is conjectured to be a tau function of a new integrable system
which can be regarded as a fractional generalization of the Volterra lattice hierarchy, so we name it the \emph{fractional Volterra hierarchy}. In this paper,
we give the definition of this integrable hierarchy in terms of Lax pair and Hamiltonian formalisms, construct its tau functions, and present its multi-soliton solutions.
\end{abstract}

\section{Introduction} \label{sec-1}

Let $g, n\in \N$ be two natural numbers satisfying the stable condition $2g-2+n>0$, and $\M_{g, n}$ be the moduli space of stable curves with genus $g$ and $n$
marked points. Denote by $\psi_k$ ($k=1, \dots, n$) the first Chern class of the $k$-th tautological line bundle $\mathcal{L}_k$ on $\M_{g, n}$, and denote by
$\lambda_j$ ($j=0, \dots, g$) the $j$-th Chern class of the Hodge bundle $\mathbb{E}_{g,n}$ on $\M_{g, n}$. The Hodge integrals are defined as the following
integrals:
\[\langle \tau_{i_1}\cdots\tau_{i_n}\lambda_{j_1}\cdots\lambda_{j_m}\rangle_{g}
:=\int_{\M_{g, n}} \psi_1^{i_1}\cdots\psi_n^{i_n}\lambda_{j_1}\cdots\lambda_{j_m}.\]
They arise naturally in the localization computation of Gromov-Witten invariants for toric varieties, and are also related to many interesting mathematical objects,
like Hurwitz numbers, irreducible representations of symmetric groups, and so on \cite{ELSV, LLZ, MV, OP}.

Denote by $\gamma_j$ the $j$-th Chern character of the Hodge bundle $\mathbb{E}_{g,n}$ on $\M_{g, n}$. They are polynomials of
$\lambda_0, \dots, \lambda_j$, and satisfy $\gamma_{2j}=0$ becuase of the Mumford relation \cite{Mum}.
The genus $g$ free energy for Hodge integrals are defined as the following generating series
\[\H_g(\mathbf{t}, \mathbf{s})=\sum_{n, m}\frac{t_{i_1}\cdots t_{i_n}}{n!}\frac{s_{j_1}\cdots s_{j_m}}{m!}
\langle \tau_{i_1}\cdots\tau_{i_n}\gamma_{2j_1-1}\cdots\gamma_{2j_m-1}\rangle_{g},\]
and the full genera free energy and the partition function are defined as
\[\H(\mathbf{t}, \mathbf{s}; \e)=\sum_{g\ge 0}\e^{2g-2} \H_g(\mathbf{t}, \mathbf{s}),
\quad \mathcal{Z}(\mathbf{t}, \mathbf{s}; \e)=e^{\H(\mathbf{t}, \mathbf{s}; \e)}. \]
We proved in \cite{DLYZ-1} that the function $w=\e^2 \p_x^2 \H$, where $\p_x=\frac{\p}{\p t_0}$, satisfies a family of partial differential equations
\[\frac{\p w}{\p t_i}=X_i\left(w, \p_x w, \p_x^2 w, \dots\right), \quad i=0, 1, 2, \dots,\]
where $X_i$ are differential polynomials, and these equations form a Hamiltonian integrable hierarchy, which is called the Hodge hierarchy. In particular,
when $s_j=0$, the Hodge hierarchy is just the Korteweg-de Vries (KdV) hierarchy, so the Hodge hierarchy is actually a deformation of the KdV hierarchy with
$s_j$'s as deformation parameters.
When $s_j$'s equal to other values, the Hodge hierarchy may coincide with some known integrable systems. For example, if
\[s_j=(2j-2)! \, s^{2j-1},\]
then we have
\[\H_g(\mathbf{t}, \mathbf{s})=\sum_{n}\frac{t_{i_1}\cdots t_{i_n}}{n!}\langle \tau_{i_1}\cdots\tau_{i_n}\P_g(s)\rangle_g,\]
where $\P_g(s)$ is the Chern polynomial of the Hodge bundle $\mathbb{E}_{g,n}$
given by
\[\P_g(s)=\lambda_0+\lambda_1\, s+\cdots+\lambda_g\, s^g.\]
So the corresponding partition function $\mathcal{Z}(\mathbf{t}, \mathbf{s}; \e)$ is also called the partition function for \emph{linear Hodge integrals}.
It is proved by Buryak in \cite{Bur-1, Bur-2} that the Hodge hierarchy for linear Hodge integrals is equivalent to the Intermediate Long Wave (ILW) hierarchy.

If we take
\[s_j=(1-4^j)\, (2j-2)! \, s^{2j-1},\]
then we have
\begin{equation}
\H_g(\mathbf{t}, \mathbf{s})=\sum_{n}\frac{t_{i_1}\cdots t_{i_n}}{n!}\langle \tau_{i_1}\cdots\tau_{i_n}
\P_g(s)\P_g(-2s)\P_g(-2s)\rangle_g. \label{cubic-hodge-1}
\end{equation}
We conjectured and proved the following theorem in \cite{DLYZ-1} and \cite{DLYZ-2} respectively:
\begin{thm}
Introduce a shift operator $\lm=e^{\e\,\p_x}$, and define
\begin{align*}
&\lm_1=\lm^{\sqrt{s}}=e^{\sqrt{s}\,\e\,\p_x},\quad \frac{\p}{\p T_k}=\sum_{i\ge0}\left(2\,k\,s\right)^i\frac{\p}{\p t_i},\\
&u=\left(\lm_1^{\frac{1}{2}}+\lm_1^{-\frac{1}{2}}\right)\left(\lm_1-2+\lm_1^{-1}\right)\H(\mathbf{t}, \mathbf{s}; \e).
\end{align*}
Then we have
\begin{align*}
2 \, \sqrt{s} \, \e \, \frac{\p u}{\p T_1}=& \left(\lm_1-\lm_1^{-1}\right)e^u,\\
12 \, \sqrt{s} \, \e \, \frac{\p u}{\p T_2}=& \left(\lm_1-\lm_1^{-1}\right)\left(e^u\left(\lm+1+\lm^{-1}\right)e^u\right),\dots.
\end{align*}
More generally, $\frac{\p u}{\p T_k}$ is given by the $k$-th equation of the Volterra lattice hierarchy (also
called the discrete KdV hierarchy),
\[k\,\binom{2k}{k}\, \sqrt{s} \, \e \, \frac{\p L}{\p T_k}=\left[\left(L^{2k}\right)_+,\ L\right],\]
where $L=\lm_1+e^u\lm_1^{-1}$.
\end{thm}

The Hodge integrals that appear in \eqref{cubic-hodge-1} belong to the class of \emph{special cubic Hodge integrals}, whose general form is given by
\begin{equation}
\langle \tau_{i_1}\cdots\tau_{i_n}\P_g(-p)\P_g(-q)\P_g(-r)\rangle_g, \label{cubic-Hodge}
\end{equation}
where $p, q, r\in \C$ satisfy the so-called \emph{local Calabi-Yau condition}
\begin{equation}
\frac{1}{p}+\frac{1}{q}+\frac{1}{r}=0. \label{local-CY}
\end{equation}
The special cubic Hodge integrals arise in the localization computation of Gromov-Witten invariants for toric Calabi-Yau threefolds, so they are very
important in the study of mirror symmetry and other related areas. Therefore, the analogue of the above theorem for more general special cubic Hodge integrals
is an important problem.

Let us take
\begin{equation}
s_j=-(2j-2)! \, \left(p^{2j-1}+q^{2j-1}+r^{2j-1}\right),\quad \mbox{where } r=-\frac{p\,q}{p+q}. \label{sj-general}
\end{equation}
Then we have
\[\H_g(\mathbf{t}, \mathbf{s})=\sum_{n}\frac{t_{i_1}\cdots t_{i_n}}{n!}\langle \tau_{i_1}\cdots\tau_{i_n}
\P_g(-p)\P_g(-q)\P_g(-r)\rangle_g.\]
Let us denote
\[\alpha=\frac{p}{\sqrt{p+q}}, \quad \beta=\frac{q}{\sqrt{p+q}},\]
and define
\begin{align}
& \lm_1=\lm^\alpha,  \quad \lm_2=\lm^\beta, \label{dfn-lm1lm2}\\
& \frac{\p}{\p x_k}=\sum_{i\ge0}\left(k\,p\right)^i\frac{\p}{\p t_i}, \quad \frac{\p}{\p y_k}=\sum_{i\ge0}\left(k\,q\right)^i\frac{\p}{\p t_i},\label{dfn-xkyk}\\
& u=\left(\lm_1\lm_2-1\right)\left(1-\lm_1^{-1}\right)\H(\mathbf{t}, \mathbf{s}; \e). \label{dfn-u}
\end{align}
Boris Dubrovin and Di Yang showed in \cite{DY} that the function $u$ satisfies, at the approximation up to $\e^{12}$, the following equation 
\begin{align}
c_1\,\e\,\frac{\p u}{\p x_1}=&\frac{\left(\lm_1\lm_2-1\right)\left(1-\lm_1^{-1}\right)}{\lm_2-1}e^u,\quad c_1=\frac{p\sqrt{p+q}}{q}.\label{new-hie-1}
\end{align}
They found that this equation is a certain reduction of  the semi-discrete 2D Toda lattice \cite{HZL, IH, S}, and they conjectured that $u$ should satisfy a certain reduction of the integrable hierarchy 
associated to the semi-discrete 2D Toda lattice.
By continuing Dubrovin-Yang's calculation, it can be shown that 
at the approximation up to $\e^{12}$ the function $u$ also satisfies the following equation 
\begin{align}
c_2\,\e\,\frac{\p u}{\p x_2}=&\frac{\left(\lm_1\lm_2-1\right)\left(1-\lm_1^{-1}\right)}{\lm_2-1}
\left(e^u\frac{\lm_1\lm_2-\lm_1^{-1}}{\lm_2-1}e^u\right) \label{new-hie-2},
\end{align}
where $c_2$ is given by
\[c_2=\frac{2p(2p+q)\sqrt{p+q}}{q^2}.\]

Note that the roles played by $p$ and $q$ are symmetric, so, if we introduce
\[\tilde{u}=\left(\lm_1\lm_2-1\right)\left(1-\lm_2^{-1}\right)\H(\mathbf{t}, \mathbf{s}; \e)=\frac{1-\lm_2^{-1}}{1-\lm_1^{-1}}u,\]
and exchange $p$ and $q$, we obtain
\[\frac{q\sqrt{p+q}}{p}\,\e\,\frac{\p \tilde{u}}{\p y_1}=\frac{\left(\lm_1\lm_2-1\right)\left(1-\lm_2^{-1}\right)}{\lm_1-1}e^{\tilde{u}},\]
which implies that
\begin{equation}
\frac{q\sqrt{p+q}}{p}\,\e\,\frac{\p u}{\p y_1}=\left(\lm_2-\lm_1^{-1}\right)e^{\frac{1-\lm_2^{-1}}{1-\lm_1^{-1}}u}.\label{new-hie-3}
\end{equation}
One can also obtain $\frac{\p u}{\p y_2}$ by using a similar argument.

The purpose of the present paper is to give a precise statement of Dubrovin-Yang's conjecture on the integrable hierarchy that governs the cubic Hodge integrals \eqref{cubic-Hodge} satisfying the local Calabi-Yau condition \eqref{local-CY}.
We will give the definition of the integrable hierarchy in terms of Lax pair representation, study its Hamiltonian
structure and tau structure, and find multi-soliton solutions.

Suppose $\alpha, \beta\in \C$ are two arbitrary complex numbers. We consider the following Lax operator:
\[L=\lm_2+e^{u(x)}\,\lm_1^{-1},\]
where $\lm_1=\lm^\alpha$, $\lm_2=\lm^\beta$, and $\lm=e^{\e\,\p_x}$.

\begin{lem}\label{lem-trivial}
Denote $\lm_3=\lm_1\lm_2$, then the following statements hold true:
\begin{itemize}
\item[i)] There exists an operator
\[A_1=\lm_3+\sum_{i\ge 0}a_{1,i}\,\left(\lm_3\right)^{-i},\]
where $a_{1,i}$ are differential polynomials of $u$, such that $A_1=L^{\frac{\alpha+\beta}{\beta}}$.
\item[ii)] There exists an operator
\[A_2=a_{2,-1}\left(\lm_3\right)^{-1}+\sum_{i\ge 0}a_{2,i}\,\left(\lm_3\right)^i,\]
where $a_{2,i}$ are differential polynomials of $u$, such that $A_2=L^{\frac{\alpha+\beta}{\alpha}}$.
\end{itemize}
\end{lem}
The definition of the fractional power of $L$ will be given in Sec.\,\ref{sec-2}.
 
\begin{dfn}
The fractional Volterra hierarchy (FVH) is defined by the following Lax equations:
\begin{equation}
\e \frac{\p L}{\p t^{i, k}}=\left[B_{i,k},\ L\right], \label{FVH}
\end{equation}
where $i=1, 2$, $k=1, 2, \dots$, and the operator $B_{i, k}$ reads
\begin{equation}
B_{i,k}=\left\{\begin{array}{rr}\left(A_1^k\right)_+, & \mbox{if } i=1; \\ -\left(A_2^k\right)_-, & \mbox{if } i=2.\end{array}\right. \label{Bik}
\end{equation}
\end{dfn}
In the above definition, the positive or negative part of an operator is given by
\[\left(\sum_{i\in\Z}a_i \left(\lm_3\right)^i\right)_+=\sum_{i\ge 0}a_i \left(\lm_3\right)^i, \quad
\left(\sum_{i\in\Z}a_i \left(\lm_3\right)^i\right)_-=\sum_{i< 0}a_i \left(\lm_3\right)^i.\]

Define the residue operation as follow
\[\Res\left(\sum_{i\in\Z}a_i \left(\lm_3\right)^i\right)=a_0.\]
Then the main results of the present paper is the following theorem.
\begin{thm}\label{thm-main}
The FVH has the following properties:
\begin{itemize}
\item[i)] It is integrable, that is, for any $(i, k)$ and $(j, l)$, we have
\[\left[\frac{\p}{\p t^{i,k}}, \frac{\p}{\p t^{j,l}}\right]=0.\]
\item[ii)] It can be written as a hierarchy of Hamiltonian systems
\[\e\,\frac{\p u}{\p t^{i,k}}=\{u, H_{i,k}\}_P,\]
where the Hamiltonian operator $P$ reads
\[P=\frac{\left(\lm_3-1\right)\left(1-\lm_1^{-1}\right)}{\lm_2-1},\]
and the Hamiltonians $H_{i,k}$ are given by
\[H_{i,k}=\frac{(i-1)\alpha+(2-i)\beta}{\left(\alpha+\beta\right)k}\int \Res \left(A_i^k\right) \d x.\]
\item[iii)] It has a tau structure. More presicely, define
\begin{align*}
\Omega_{1, k; j, l}=\sum_{m=1}^k \frac{\lm_3^m-1}{\lm_3-1}\left(\Res\left(\lm_3^{-m}A_1^k\right)\Res\left(A_j^l\lm_3^{m}\right)\right),\\
\Omega_{2, k; j, l}=\sum_{m=1}^k \frac{\lm_3^m-1}{\lm_3-1}\left(\Res\left(A_2^k\lm_3^{m}\right)\Res\left(\lm_3^{-m}A_j^l\right)\right),
\end{align*}
then for any solution $u(x; \mathbf{t})$, there exists a function $\tau(x; \mathbf{t})$, such that
\begin{align}
& \left(\lm_3-1\right)\left(1-\lm_1^{-1}\right)\log \tau=u, \label{tau-dfn-1}\\
&\e\left(\lm_3-1\right)\frac{\p \log \tau}{\p t^{i, k}}=\Res A_i^k, \label{tau-dfn-2}\\
&\e^2 \frac{\p^2 \log \tau}{\p t^{i,k} \p t^{j,l}}=\Omega_{i, k; j, l}.\label{tau-dfn-3}
\end{align}
\end{itemize}
\end{thm}

The Dubrovin-Yang conjecture for the special cubic Hodge integrals can be stated as the following \emph{Hodge-FVH correspondence}.
\begin{cnj}[Hodge-FVH correspondence]\label{main-conj}
Let $\H(\mathbf{t}, \mathbf{s}; \e)$ be the free energy of Hodge integrals with $s_j$ given in \eqref{sj-general},
and $\frac{\p}{\p x_k}$, $\frac{\p}{\p y_k}$, $u$ be given in \eqref{dfn-xkyk}, \eqref{dfn-u}.
Then the recombined Hodge hierarchy
\[\left\{\frac{\p u}{\p x_k}=\sum_{i\ge0}\left(k\,p\right)^i\frac{\p u}{\p t_i},
\quad \frac{\p u}{\p y_k}=\sum_{i\ge0}\left(k\,q\right)^i\frac{\p u}{\p t_i}\right\}\]
coincides with the FVH \eqref{FVH} after the rescaling
\[x_k=\alpha\,k\,\binom{\frac{\alpha+\beta}{\beta}k}{k}\,t^{1, k}, \quad y_k=\beta\,k\,\binom{\frac{\alpha+\beta}{\alpha}k}{k}\,t^{2, k}.\]
\end{cnj}

The paper is arranged as follow: In Sec.\,\ref{sec-2} we give the definition of the Lax operator $L$ and its fractional powers. In Sec.\,\ref{sec-3} we prove the main theorem of the paper. In Sec.\,\ref{sec-4} we present the multi-soliton solutions of the FVH. Finally, in Sec.\,\ref{sec-5} we give some concluding remarks. 

\section{Fractional pseudo-difference operators}\label{sec-2}

Let $\A$ be a commutative $\C$-algebra with a gradation
\[\A=\prod_{i\ge 0}\A_i,\quad \A_i\cdot \A_j\subset \A_{i+j},\]
such that $\A$ is topological complete with respect to the induced decreasing filtration
\[\A=\A_{(0)}\supset \cdots \supset \A_{(d-1)}\supset \A_{(d)}\supset \A_{(d+1)}\supset \cdots,\quad \A_{(d)}=\prod_{i\ge d}\A_i.\]
An element $f\in\A$ can be decomposed as
\[f=f_0+f_1+f_2+\cdots, \quad f_i\in\A_i.\]
We often write it as
\[f=f_0+\e f_1+\e^2 f_2+\cdots,\]
where $\e$ is just a formal parameter, whose power counts the degree.

Let $\p_x:\A\to\A$ be a derivation with degree one, that is $\p_x\A_i\subset\A_{i+1}$.
Since $\A$ is complete, the exponential map of $\p_x$ is well-defined, which is called the shift operator, and is denoted by
\[\lm=e^{\e\,\p_x}=\sum_{k\ge 0} \frac{1}{k!}\e^k \p_x^k.\]
Then, for any complex number $\mu\in\C$, we can defined the $\mu$-th power of $\lm$ as follow:
\[\lm^\mu=e^{\e\,\mu\,\p_x}=\sum_{k\ge 0} \frac{1}{k!}\e^k \mu^k \p_x^k.\]

Let $\Delta\in\C$ be a complex number. The ring of fractional pseudo-difference operators with step size $\Delta$ is defined as the following $\A$-module
\[\D(\Delta)=\left\{\left.\sum_{k\ge 0} f_k \,\lm^{\gamma-k \Delta}\right| \gamma\in\C,\ f_i \in \A \right\},\]
equipped with the following multiplication:
\begin{align*}
&\left(\sum_{k\ge 0} f_k \,\lm^{\gamma_1-k \Delta}\right)\cdot\left(\sum_{l\ge 0} g_l \,\lm^{\gamma_2-l \Delta}\right)\\
=&\sum_{k\ge 0}\left(\sum_{i=0}^k f_i \lm^{\gamma_1-i\Delta}\left(g_{k-i}\right)\right)\lm^{\gamma_1+\gamma_2-k\Delta}.
\end{align*}
We adopt the $\lm^{-\Delta}$-adic topology on $\D(\Delta)$, so that its elements are convergent.

We further assume that there is a suitable topology on $\A_0$ such that the exponential map
\[\exp: \A_0\to\A_0, \quad u\mapsto e^u=\sum_{k\ge 0} \frac{1}{k!}u^k\]
is well-defined. Then it is easy to see that $\exp:\A\to\A$ is also well-defined, since
\[e^{f_0+f_1+f_2+\cdots}=e^{f_0}\cdot e^{f_1+f_2+\cdots},\]
where $e^{f_1+f_2+\cdots}$ does exist due to the completeness of $\A$.

Suppose $u\in \A$, and $L\in \D(\Delta)$ has the following form
\begin{equation}
L=e^u \lm^\gamma+f_1 \lm^{\gamma-\Delta}+f_2 \lm^{\gamma-2\Delta}+\cdots, \quad \gamma\ne 0. \label{Leu}
\end{equation}
The main goal of the this section is to construct the $\mu$-th power of $L$ for an arbitrary $\mu \in \C$.

We consider the case with $L=e^u \lm^\gamma$ first. By using the integral form of the Baker-Campbell-Hausdorff (BCH) formula (see e.g. \cite{Hall}),
one can obtain the following identity:
\begin{equation}
\log \left(e^u\,\lm^\gamma\right)=\gamma\,\e\,\p_x+\frac{\gamma\,\e\,\p_x}{e^{\gamma\,\e\,\p_x}-1}\left(u\right),
\end{equation}
so the $\mu$-th power of $L=e^u \lm^\gamma$ can be defined as
\begin{align*}
\left(e^u\,\lm^\gamma\right)^\mu=&\exp\left(\mu\,\gamma\,\e\,\p_x+\frac{\mu\,\gamma\,\e\,\p_x}{e^{\gamma\,\e\,\p_x}-1}\left(u\right)\right)\\
=&\exp\left(\mu\,\gamma\,\e\,\p_x+\frac{\mu\,\gamma\,\e\,\p_x}{e^{\mu\,\gamma\,\e\,\p_x}-1}
\frac{e^{\mu\,\gamma\,\e\,\p_x}-1}{e^{\gamma\,\e\,\p_x}-1}\left(u\right)\right)\\
=&\exp\left(\frac{\lm^{\mu\,\gamma}-1}{\lm^\gamma-1}\left(u\right)\right)\lm^{\mu\,\gamma}.
\end{align*}

On the other hand, if
\[L=1+f_1 \lm^{-\Delta}+f_2 \lm^{-2\Delta}+\cdots,\]
its logrithm is given by
\[\log\left(L\right)=\sum_{k\ge1}\frac{(-1)^{k-1}}{k}\left(f_1 \lm^{-\Delta}+f_2 \lm^{-2\Delta}+\cdots\right)^k,\]
which is convergent in the $\lm^{-\Delta}$-adic topology of $\D(\Delta)$.

Finally, if $L$ is given by \eqref{Leu}, we can write it as $L=L_1L_2$, where
\[L_1=e^u\,\lm^\gamma, \quad L_2=1+e^{-u}\,\lm^{-\gamma}(f_1)\lm^{-\Delta}+e^{-u}\,\lm^{-\gamma}(f_2)\lm^{-2\Delta}+\cdots.\]
We define $P=\log L_1$, $Q=\log L_2$, then
\[\log L=\log \left(\exp P\,\exp Q\right),\]
which can be obtained from the BCH formula again. Then the $\mu$-th power of $L$ is defined as
\[L^\mu=\exp\left(\mu \log L\right).\]

\section{Proofs of the main results}\label{sec-3}

Let $u$ be the coordinate on $\C$, and $\A_0$ be the $\C$-algebra of entire functions on $\C$. We define
\[\A=\A_0[[u^i\mid i=1, 2, \dots]],\]
which is completed with respect to the following degree
\[\deg f(u)=0, \quad \deg u_i=i.\]
Then $\A$ satisfies the assumptions given in the last section. Define
\[\p_x=\sum_{i\ge 0}u^{i+1}\frac{\p}{\p u^i},\]
then $\p_x$ is a degree-one derivation, so we can define its exponential map $\lm$ on $\A$, and the ring $\D(\Delta)$ of
fractional pseudo-difference operators with an arbitrary step size $\Delta\in\C$.

Let $\alpha, \beta\in \C$ be two arbitrary complex numbers, and $\Delta=\alpha+\beta$, then the following operator
\[L=\lm^\beta+e^u\,\lm^{-\alpha}\]
is an element of $\D(\Delta)\cap \D(-\Delta)$, which is called the Lax operator of the FVH.

According to the definitions given in the last section, we can define the $\mu$-th power of $L$ in both $\D(\Delta)$ and $\D(-\Delta)$
for an arbitrary $\mu\in\C$. We denote
\[A_1=L^{\frac{\alpha+\beta}{\beta}}\in \D(\Delta), \quad A_2=L^{\frac{\alpha+\beta}{\alpha}}\in \D(-\Delta),\]
then they fulfill the requirements asked in Lemma \ref{lem-trivial}, so this lemma is proved trivially. Then we can define $B_{i,k}$ as \eqref{Bik},
and define the FVH by \eqref{FVH}.

It is needed to explain that why the equation \eqref{FVH} gives a partial differential equation for $u$. We take the $i=1$ case for example.
Denote by $B=\left(A_1^k\right)_+$, then there is $B'=\left(A_1^k\right)_-$ such that
\[[B+B', L]=[L^{\frac{\alpha+\beta}{\beta}k}, L]=0,\]
so the equation \eqref{FVH} can be written as
\[\e\,\frac{\p L}{\p t}=[B, L]=-[B', L].\]
Let $L'=L\lm_1=\lm_3+e^u$, then we have
\[\e\,e^u\,\frac{\p u}{\p t}=\e\,\frac{\p L'}{\p t}=B L'-L'\left(\lm_1^{-1}B\lm_1\right)=-B' L'+L'\left(\lm_1^{-1}B'\lm_1\right).\]
In the above equation, both of the second term and the third term are formal power series of $\lm_3$.
However, the second term doesn't contain negative powers of $\lm_3$, while the third term doesn't contain positive powers of $\lm_3$,
so both of them should be equal to the zeroth power of $\lm_3$, so it gives a partial differential equation for $u$. The $i=2$ case is similar.

\begin{emp}
According to the definitions given in the last section, one can show that
\[A_1=\lm_3+\frac{\lm_3-1}{\lm_2-1}\left(e^u\right)+O\left(\lm_3^{-1}\right),\]
then it is easy to obtain the $t^{1,1}$-flow of the FVH
\[\e\frac{\p u}{\p t^{1,1}}=\frac{\left(\lm_3-1\right)\left(1-\lm_1^{-1}\right)}{\lm_2-1}\left(e^u\right).\]
\end{emp}

\begin{emp}
Similarly, we have
\[A_2=e^{\frac{1-\lm_3^{-1}}{1-\lm_1^{-1}}u}\lm_3^{-1}+\frac{\lm_2-\lm_1^{-1}}{1-\lm_1^{-1}}e^{\frac{1-\lm_2^{-1}}{1-\lm_1^{-1}}u}
+O\left(\lm_3\right),\]
so the $t^{2,1}$-flow of the FVH reads
\[\e\frac{\p u}{\p t^{2,1}}=\left(\lm_2-\lm_1^{-1}\right)e^{\frac{1-\lm_2^{-1}}{1-\lm_1^{-1}}u}.\]
\end{emp}

\begin{rmk}
The local Calabi-Yau condition \eqref{local-CY} is symmetric in $p$, $q$, $r$, so there should exist an $S_3$-action on $\H(\mathbf{t}, \mathbf{s})$.
In Sect.\,\ref{sec-1}, we used the reflection $\sigma_{12}: p\leftrightarrow q \in S_3$ to obtain the flow $\frac{\p}{\p y_1}$ from the flow $\frac{\p}{\p x_1}$.
The other elements of $S_3$ can be also used. For example, if we intoduce the following flows:
\[\frac{\p}{\p z_k}=\sum_{i\ge0}\left(k\,r\right)^i\frac{\p}{\p t_i},\]
and consider the action of $\sigma_{13}: p\leftrightarrow r$ or  $\sigma_{23}: q\leftrightarrow r$ on $\H(\mathbf{t}, \mathbf{s})$, then one can obtain
the flow $\frac{\p}{\p z_1}$ from $\frac{\p}{\p x_1}$ or $\frac{\p}{\p y_1}$:
\[\frac{p q}{\left(p+q\right)\sqrt{p+q}} \, \e \, \frac{\p u}{\p z_1}=\left(\lm_1^{-1}-1\right)e^{-\frac{\lm_2-1}{\lm_2-\lm_1^{-1}}u}.\]
One can also find other flows $\frac{\p}{\p z_k}$ by using the similar method.
However, we do not know how to represent these new flows in the Lax form \eqref{FVH}.
\end{rmk}

The following lemma can be easily proved by using the definition \eqref{FVH}, \eqref{Bik} of the 
FVH, so we omit its proof.
\begin{lem}
For $i, j=1, 2$, $k,l=1, 2, \dots$, we have
\begin{equation}
\e \frac{\p B_{j,l}}{\p t^{i,k}}-\e \frac{\p B_{i,k}}{\p t^{j,l}}+[B_{j,l}, B_{i,k}]=0.
\end{equation}
\end{lem}
By using this lemma, one can proved the first part of Theorem \ref{thm-main}, that is, the equation \eqref{FVH} indeed define an integrable system.
To prove the second part of the theorem, we need to use some identities about the variational derivatives of $H_{i,k}$.

\begin{lem}
For $i=1, 2$, and $k=1, 2, \dots$, we have
\begin{equation}
\frac{\delta H_{i, k}}{\delta u}=\Res\left(e^u \, \lm_1^{-1}\, A_i^k\, L^{-1}\right), \label{var-der}\\
\end{equation}
\end{lem}
\begin{prf}
We prove the $i=1$ case only, since the proof for the $i=2$ case is similar. Recall that
\[A_1^{k}=L^{\frac{\alpha+\beta}{\beta}\, k}=\exp\left(\frac{\alpha+\beta}{\beta}\, k \, \log L\right).\]
By using the following well-known identity
\[\int \Res \left(X Y\right)\, \d x=\int \Res \left(Y X\right)\, \d x,\]
one can show that
\[\int \Res e^{A+\delta A}\, \d x=\int \Res e^A\left(1+\delta A\right)\, \d x+O(\delta A^2),\]
so we have
\begin{align*}
&\delta \int \Res A_1^k \, \d x=\delta \int \Res \exp\left(\frac{\alpha+\beta}{\beta}\, k \, \log L\right) \, \d x \\
=& \frac{\alpha+\beta}{\beta}\, k \, \int \Res A_1^k \, \left(\delta \log L\right)\, \d x+O(\left(\delta \log L\right)^2).
\end{align*}
On the other hand, we write $L(u+\delta u)$ as
\[L(u+\delta u)=L+\delta L+O(\delta u^2)=e^{\log L} e^{L^{-1}\delta L+O(\delta u^2)},\]
then, by using the BCH formula, we obtain
\begin{align*}
&\log L(u+\delta u)=\log \left(e^{\log L} e^{L^{-1}\delta L+O(\delta u^2)}\right)\\
=& \log L+ \frac{\ad_{\log L}}{1-e^{-\ad_{\log L}}}\left(L^{-1}\delta L\right)+O(\delta u^2).
\end{align*}
So we have
\begin{align*}
&\int \Res A_1^k \, \left(\delta \log L\right)\, \d x\\
=&\int \Res  \frac{\ad_{\log L}}{1-e^{-\ad_{\log L}}}\left(L^{-1}\delta L\right) A_1^k\, \d x+O(\delta u^2)\\
=& \int \Res  L^{-1}\delta L\, \frac{\ad_{\log L}}{e^{\ad_{\log L}}-1} \left(A_1^k\right)\, \d x+O(\delta u^2)\\
=&\int \Res  \delta L \, A_1^k\, L^{-1}\, \d x+O(\delta u^2),
\end{align*}
where we used the following identity:
\[\int \Res \left(\ad_X(Y) Z\right)\, \d x=-\int \Res \left(Y \ad_X(Z)\right)\, \d x.\]
Finally,  by replacing $\delta L$ by $\delta u\,e^u\, \lm^{-1}$, the lemma is proved.
\end{prf}

\begin{lem}
For $i=1, 2$, and $k=1, 2, \dots$, we have
\begin{equation}
\Res A_i^k =\frac{\lm_3-1}{\lm_2-1}\Res\left(e^u \lm_1^{-1}A_i^k L^{-1}\right)
\end{equation}
\end{lem}
\begin{prf}
Suppose in $\D(\Delta)$ or $\D(-\Delta)$ we have
\[A_i^k L^{-1}=\lm_1 e^{-u} \left(\sum_{i\in \Z} b_i \lm_3^i\right),\]
where the summation is truncated on the positive or negative direction. So we have
\[\Res\left(e^u \lm_1^{-1}A_i^k L^{-1}\right)=b_0.\]
Then we use two methods to compute $\Res A_i^k$:
\begin{align*}
R_1 =& \Res \left(A_i^k L^{-1}\right) L =\Res \lm_1 e^{-u} \left(\sum_{i\in \Z} b_i \lm_3^i\right)\left(\lm_2+e^u \lm_1^{-1}\right)\\
=& \Res \sum_{i\in\Z} \lm_1\left(e^{-u}\left(b_{i-1}+b_i\,\lm_3^i\left(e^u\right)\right)\right)\lm_3^i=\lm_1\left(b_0+e^{-u}b_{-1}\right)\\
R_2=& \Res L\left(A_i^k L^{-1}\right)=\Res \left(\lm_2+e^u \lm_1^{-1}\right)\lm_1 e^{-u} \left(\sum_{i\in \Z} b_i \lm_3^i\right)\\
=& \Res \sum_{i\in\Z} \left(b_i+\lm_3\left(e^{-u}\,b_{i-1}\right)\right)\lm_3^i=b_0+\lm_3\left(e^{-u}\,b_{-1}\right).
\end{align*}
It is easy to see that $\lm_2\left(R_1\right)-R_2=\left(\lm_3-1\right)b_0$, so we have
\[\Res A_i^k=R_1=R_2=\frac{\lm_3-1}{\lm_2-1} b_0+C_{i, k},\]
where $C_{i,k}$ belong to the kernel of $\lm_2-1$, so they must be constants. Finally, by taking $u=0$, one can show that $C_{i, k}=0$. The lemma is proved.
\end{prf}

By using the above two identities, we have
\begin{align*}
\e\,\frac{\p u}{\p t^{i,k}}=& e^{-u}\left[B_{i, k}, L\right]\lm_1=\left(1-\lm_1^{-1}\right)\Res A_i^k\\
=& \left(1-\lm_1^{-1}\right)\frac{\lm_3-1}{\lm_2-1}\Res\left(e^u \lm_1^{-1}A_i^k L^{-1}\right)=P\left(\frac{\delta H_{i,k}}{\delta u}\right).
\end{align*}
The operator $P$ is obviously a Hamiltonian operator, so the second part of Theorem \ref{thm-main} is proved.

To prove the third part of Theorem \ref{thm-main}, we first derive some identities that are satisfied by the functions $\Omega_{i, k; j, l}$ in the following lemma.

\begin{lem}
For $i, j=1, 2$, $k, l=1, 2, \dots$, we have
\begin{equation}
\e\,\frac{\p}{\p t^{i, k}}\Res A_j^l=\e\,\frac{\p}{\p t^{j, l}}\Res A_i^k=\left(\lm_3-1\right)\Omega_{i, k; j, l}.
\end{equation}
\end{lem}
\begin{prf}
We only prove the $i=1$ case, since the $i=2$ case is similar.
The first identity can be easily proved by using the definition \eqref{FVH}, \eqref{Bik} of the FVH. 
To prove the second one, we first represent $A_i^k$ and $A_j^l$ in the forms
\[A_i^k=\sum_{m\in\Z} \Res\left(A_i^k \lm_3^{-m}\right)\lm_3^m,\quad A_j^l=\sum_{n\in\Z} \Res\left(A_j^l \lm_3^{-n}\right)\lm_3^n,\]
then we have
\begin{align*}
& \e\,\frac{\p}{\p t^{i, k}}\Res A_j^l=\Res\left[\left(A_i^k\right)_+, A_j^l\right]\\
=&\Res\left[\sum_{m=0}^k \Res\left(A_i^k \lm_3^{-m}\right)\lm_3^m, \sum_{n\in\Z} \Res\left(A_j^l \lm_3^{-n}\right)\lm_3^n\right]\\
=& \sum_{m=1}^k\left(\Res\left(A_i^k \lm_3^{-m}\right)\lm_3^m\left(\Res\left(A_j^l \lm_3^{m}\right)\right)\right.\\
&\qquad\qquad \left.-\Res\left(A_j^l \lm_3^{m}\right)\lm_3^{-m}\left(\Res\left(A_i^k \lm_3^{-m}\right)\right)\right)\\
=&\sum_{m=1}^k\left(\lm_3^m-1\right)\left(\Res\left(\lm_3^{-m}A_i^k\right)\Res\left(A_j^l \lm_3^{m}\right)\right)\\
=&\left(\lm_3-1\right)\Omega_{i, k; j, l}.
\end{align*}
The lemma is proved.
\end{prf}

\begin{lem}\label{lem-iden}
The differential polynomials $\Omega_{i, k; j, l}$ satisfy the following identities:
\begin{align*}
&\Omega_{i, k; j, l}=\Omega_{j, l; i, k},\\
& \frac{\p}{\p t^{i, k}}\Omega_{j, l; m, n}=\frac{\p}{\p t^{j, l}}\Omega_{i, k; m, n},
\end{align*}
where $i, j, m=1, 2$, $k, l, n=1, 2, \dots$.
\end{lem}
\begin{prf}
The above lemma implies that
\begin{align*}
&\left(\lm_3-1\right)\left(\Omega_{i, k; j, l}-\Omega_{j, l; i, k}\right)=0,\\
& \left(\lm_3-1\right)\left(\frac{\p}{\p t^{i, k}}\Omega_{j, l; m, n}-\frac{\p}{\p t^{j, l}}\Omega_{i, k; m, n}\right)=0,
\end{align*}
so there are constants $C_{i, k; j, l}$ and $C'_{i, k; j, l; m, n}$ such that
\begin{align}
&\Omega_{i, k; j, l}-\Omega_{j, l; i, k}=C_{i, k; j, l}, \label{kernel-1}\\
& \frac{\p}{\p t^{i, k}}\Omega_{j, l; m, n}-\frac{\p}{\p t^{j, l}}\Omega_{i, k; m, n}=C'_{i, k; j, l; m, n}. \label{kernel-2}
\end{align}
Note that the terms on the left hand side of \eqref{kernel-2} have positive degrees, so we have $C'_{i, k; j, l; m, n}=0$.
To find $C_{i, k; j, l}$, let us take $u=0$ and $u^i=0$, then the binomial series gives
\[\Res \lm_3^{-m}A_1^k=\binom{\frac{\alpha+\beta}{\beta}k}{k-m}, \quad \Res \lm_3^{n}A_2^l=\binom{\frac{\alpha+\beta}{\alpha}l}{l-n}.\]
so we have
\begin{align*}
\Omega_{1, k; 1, l}=&\sum_{m=1}^k m \binom{\frac{\alpha+\beta}{\beta}k}{k-m}\binom{\frac{\alpha+\beta}{\beta}l}{l+m}
=\frac{\alpha\,k\,l}{(k+l)(\alpha+\beta)}\binom{\frac{\alpha+\beta}{\beta}k}{k}\binom{\frac{\alpha+\beta}{\beta}l}{l},\\
\Omega_{1, k; 2, l}=&\sum_{m=1}^k m \binom{\frac{\alpha+\beta}{\beta}k}{k-m}\binom{\frac{\alpha+\beta}{\alpha}l}{l-m}
=\frac{\alpha\,\beta\,k\,l}{(\alpha+\beta)(k\,\alpha+l\,\beta)}\binom{\frac{\alpha+\beta}{\beta}k}{k}\binom{\frac{\alpha+\beta}{\alpha}l}{l},\\
\Omega_{2, k; 1, l}=&\sum_{m=1}^k m \binom{\frac{\alpha+\beta}{\alpha}k}{k-m}\binom{\frac{\alpha+\beta}{\beta}l}{l-m}
=\frac{\alpha\,\beta\,k\,l}{(\alpha+\beta)(k\,\beta+l\,\alpha)}\binom{\frac{\alpha+\beta}{\alpha}k}{k}\binom{\frac{\alpha+\beta}{\beta}l}{l},\\
\Omega_{2, k; 2, l}=&\sum_{m=1}^k m \binom{\frac{\alpha+\beta}{\alpha}k}{k-m}\binom{\frac{\alpha+\beta}{\alpha}l}{l+m}
=\frac{\beta\,k\,l}{(k+l)(\alpha+\beta)}\binom{\frac{\alpha+\beta}{\alpha}k}{k}\binom{\frac{\alpha+\beta}{\alpha}l}{l},
\end{align*}
which imply $C_{i, k; j, l}=0$. A proof of these identities and their generalization can be found in \cite{DP}. The lemma is proved.
\end{prf}

Finally, to prove the last part of Theorem \ref{thm-main}, we only need to verify the compatibility of the equations \eqref{tau-dfn-1}-\eqref{tau-dfn-3},
which has been proved in the above two lemmas. So Theorem \ref{thm-main} is now proved.

\section{Darboux transformations and multi-soliton solutions}\label{sec-4}

The Lax equations \eqref{FVH} of the FVH 
are the compatibility conditions of the following linear systems: 
 \[L\psi =\lambda \psi,\quad \e\frac{\p \psi}{\p t^{i, k}}=B_{i, k}\psi.\]
In this section, we are to apply Darboux transformation method \cite{GHZ, Matv} to these linear systems
in order to obtain special solutions, in particular soliton solutions, of the FVH. 

Suppose $u(x; \mathbf{t})$ is a solution of the FVH, and $\lambda$ is an eigenvalue of the corresponding Lax operator $L$,
then one can find a solution $\psi(x; \mathbf{t})$ of the above Lax pair, which is called a wave function for the potential $u$ and the eigenvalue $\lambda$.

\begin{lem}[Darboux transformation]
Suppose $\psi$ is a wave function for the potential $u$ and the eigenvalue $\lambda$. Define operators
\begin{equation}
W=\lm_3-\frac{\lm_3\left(\psi\right)}{\psi}, \quad \tilde{L}=W L W^{-1}, \label{darboux-1}
\end{equation}
then $\tilde{L}$ has the form $\tilde{L}=\lm_2+e^{\tilde{u}}\lm_1^{-1}$, where
\begin{equation}
\tilde{u}=u+\left(\lm_3-1\right)\left(1-\lm_1^{-1}\right)\log\psi. \label{darboux-2}
\end{equation}
Moreover, $\tilde{u}$ is also a solution of the FVH.
\end{lem}
\begin{prf}
It is easy to verify the expression \eqref{darboux-2}. To prove that $\tilde{u}$ is also a solution of the FVH, one need to show
\[\e \frac{\p \tilde{L}}{\p t^{i,k}}=\left[\tilde{B}_{i, k},\, \tilde{L}\right],\]
which is implied by the equation
\begin{equation}
\e \frac{\p W}{\p t^{i, k}}=\tilde{B}_{i, k} W-W B_{i, k}. \label{eq-wt}
\end{equation}
Here $\tilde{B}_{i, k}$ is the operator \eqref{Bik} with $L$ replaced by $\tilde{L}$.

We take the $i=1$ case for example. Then
\[B_{i, k}=\left(A_1^k\right)_+, \quad \tilde{B}_{i, k}=\left(W A_1^k W^{-1}\right)_+.\]
It is easy to see that
\[\left(W A_1^k W^{-1}\right)_+ W-W \left(A_1^k\right)_+ = W \left(A_1^k\right)_- - \left(W A_1^k W^{-1}\right)_- W,\]
so the equation \eqref{eq-wt} also can be written as
\begin{equation}
\e \frac{\p}{\p t^{i, k}}\left(-\frac{\lm_3\left(\psi\right)}{\psi}\right)=\Res \left( W A_1^k W^{-1}- A_1^k\right)
\left(-\frac{\lm_3\left(\psi\right)}{\psi}\right). \label{eq-psipsi}
\end{equation}
By using the following expression of the operator $W^{-1}$ (in $\D(\Delta)$):
\[W^{-1}=\sum_{m\ge0}\frac{\psi}{\lm_3^{-m}\left(\psi\right)}\lm_3^{-m-1},\]
one can show that
\[\Res \left( W A_1^k W^{-1}- A_1^k\right)=\left(\lm_3-1\right)\left(\frac{\left(A_1^k\right)_+\psi}{\psi}\right),\]
which implies \eqref{eq-psipsi} immediately. The lemma is proved.
\end{prf}

By applying the Darboux transformation $n$ times, we obtain the following theorem. Its proof is standard \cite{GHZ, Matv}, so we omit it here.
\begin{thm}[$n$-fold Darboux transformation]
Suppose $u$ is a solution of the FVH, and $\psi_i\ (i=1, \dots, n)$ are the wave functions for the eigenvalues $\lambda_i\ (i=1, \dots, n)$ of $L$.
Define operators
\begin{equation}
W=\frac{E_n}{D_n}, \quad \tilde{L}=W L W^{-1}, \label{darboux-n1}
\end{equation}
where
\begin{align}
E_n=&\left|\begin{array}{cccc}
\lm_3^n & \lm_3^{n-1} & \cdots & \lm_3^0\\
\lm_3^n\left(\psi_1\right) & \lm_3^{n-1}\left(\psi_1\right) & \cdots & \lm_3^0\left(\psi_1\right)\\
\vdots & \vdots & \ddots & \vdots\\
\lm_3^n\left(\psi_n\right) & \lm_3^{n-1}\left(\psi_n\right) & \cdots & \lm_3^0\left(\psi_n\right)
\end{array}\right|, \label{formula-En}\\ & \nonumber\\
D_n=&\left|\begin{array}{cccc}
\lm_3^{n-1}\left(\psi_1\right) & \lm_3^{n-2}\left(\psi_1\right) & \cdots & \lm_3^0\left(\psi_1\right)\\
\lm_3^{n-1}\left(\psi_2\right) & \lm_3^{n-2}\left(\psi_2\right) & \cdots & \lm_3^0\left(\psi_2\right)\\
\vdots & \vdots & \ddots & \vdots\\
\lm_3^{n-1}\left(\psi_n\right) & \lm_3^{n-2}\left(\psi_n\right) & \cdots & \lm_3^0\left(\psi_n\right)
\end{array}\right|, \label{formula-Dn}
\end{align}
then $\tilde{L}$ has the form $\tilde{L}=\lm_2+e^{\tilde{u}}\lm_1^{-1}$, where
\begin{equation}
\tilde{u}=u+\left(\lm_3-1\right)\left(1-\lm_1^{-1}\right)\log D_n. \label{darboux-n2}
\end{equation}
Moreover, $\tilde{u}$ is also a solution of the FVH.
\end{thm}

It is easy to see that the function $u(x; \mathbf{t})=0$ is a trivial solution of FVH. Starting from this solution, we can obtain the $n$-soliton solutions of
FVH by using the above theorem.

Denote by $L_0=\lm_2+\lm_1^{-1}$, and assume that $\lambda\in\C$ is an eigenvalue of $L_0$.
Suppose $z$ is a root of the equation $z^\beta+z^{-\alpha}=\lambda$, then it is easy to see that the function
\[\psi(x; \mathbf{t}; z)=\exp\left(\frac{x}{\e}\log z+\frac{1}{\e}\sum_{i=1}^2 \sum_{k\ge 1}b_{i,k}(z)t^{i, k}\right),\]
where
\[
b_{1, k}(z)= \sum_{m=0}^k \binom{\frac{\alpha+\beta}{\beta}k}{k-m}z^{(\alpha+\beta)m},\quad
b_{2, k}(z)=-\sum_{m=1}^k \binom{\frac{\alpha+\beta}{\alpha}k}{k-m}z^{-(\alpha+\beta)m},
\]
is a wave function for the eigenvalue $\lambda$.

We assume $\alpha, \beta$ are two positive real numbers from now on, and $\lambda$ is a real number such that
\begin{equation}
\lambda > \left(\frac{\alpha}{\beta}\right)^{\frac{\beta}{\alpha+\beta}}+\left(\frac{\beta}{\alpha}\right)^{\frac{\alpha}{\alpha+\beta}},
\label{cond-lm}
\end{equation}
then the equation $z^\beta+z^{-\alpha}=\lambda$ always has two real roots $z_1(\lambda)$ and $z_2(\lambda)$, such that
\[z_1(\lambda)<\left(\frac{\alpha}{\beta}\right)^{\frac{1}{\alpha+\beta}}<z_2(\lambda).\]
In this case, the general real wave function has the following form:
\[\psi(\lambda)=C_1 \,\psi(x; \mathbf{t}; z_1(\lambda))+C_2 \, \psi(x; \mathbf{t}; z_2(\lambda)),\]
where $C_i\ (i=1, 2)$ are arbitrary real numbers.

To construct the $n$-soliton solution, let us take $\lambda_1, \dots, \lambda_n\in\R$ satisfying the condition \eqref{cond-lm}, and denote
\[\psi_m=C_{1,m} \,\psi(x; \mathbf{t}; z_1(\lambda_m))+C_{2,m} \, \psi(x; \mathbf{t}; z_2(\lambda_m)),\quad m=1, \dots, n,\]
where $C_{i, m}\ (i=1, 2;\ m=1, \dots, n)$ are some real numbers.
Then we have the $n$-soliton solution
\[u=\left(\lm_3-1\right)\left(1-\lm_1^{-1}\right)\log D_n,\]
where $D_n$ is given by \eqref{formula-Dn}. This expression also shows that $D_n$ is the tau function for the solution
$u$. Note that not all choices of $C_{i, m}$ lead to a positive $D_n$.
It is an interesting problem to find the condition such that $D_n>0$ for all $x, \mathbf{t}\in \R$.

\begin{emp}
Take $\alpha=1.1$, $\beta=1.9$,
\[\lambda_1=3.2,\quad \lambda_2=3.8, \quad \lambda_3=4.2,\quad \lambda_4=5.2,\]
and $C_{i,m}=(-1)^{(i-1)(m-1)}$ for $i=1, 2$,  $m=1, \dots, 4$. Then the function $u(x, t^{1,1})$ gives a $4$-soliton solution for the $t^{1,1}$-flow of the FVH,
see Figure \ref{fig-1}-\ref{fig-4}.
\begin{figure}
\begin{minipage}[t]{0.5\linewidth}
\centering
\includegraphics[width=0.8\textwidth]{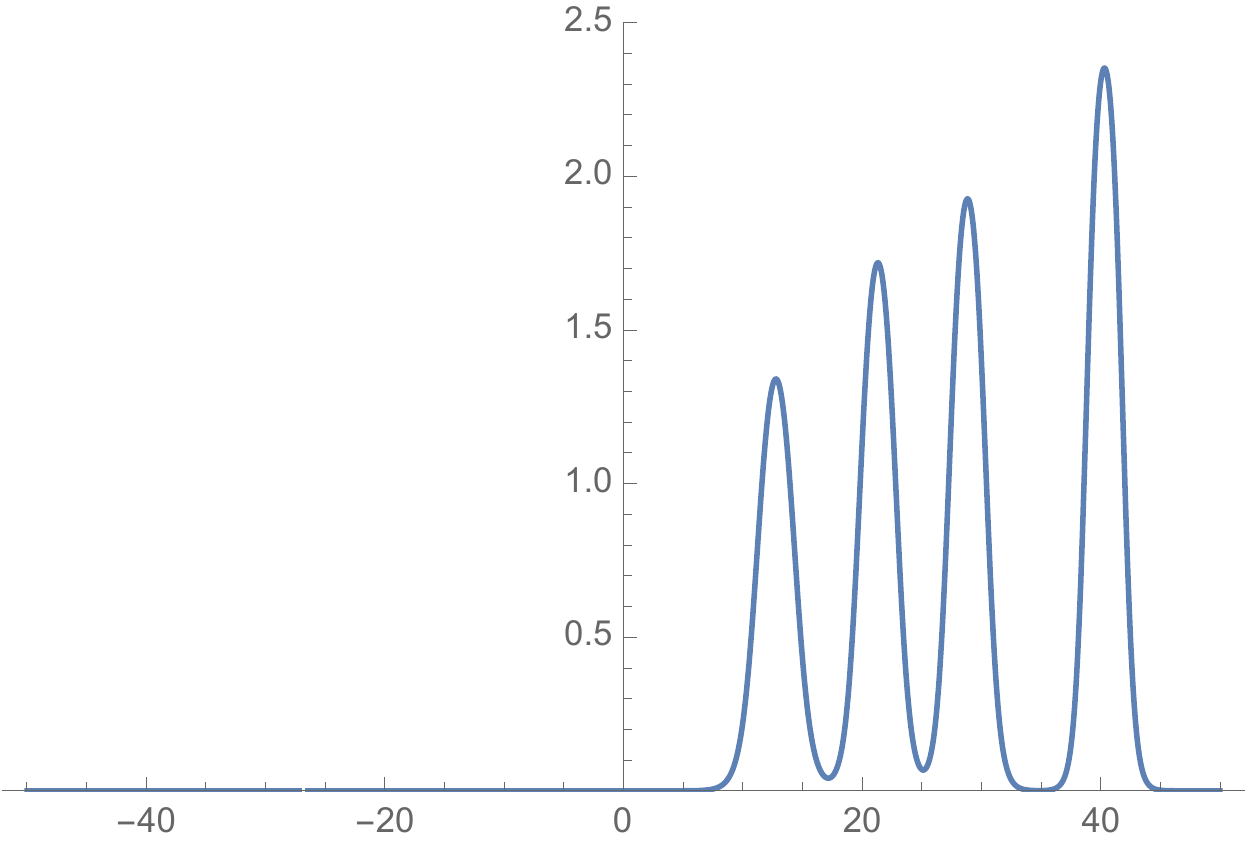}
\caption{$t^{1,1}=-8$}\label{fig-1}
\end{minipage}
\begin{minipage}[t]{0.5\linewidth}
\centering
\includegraphics[width=0.8\textwidth]{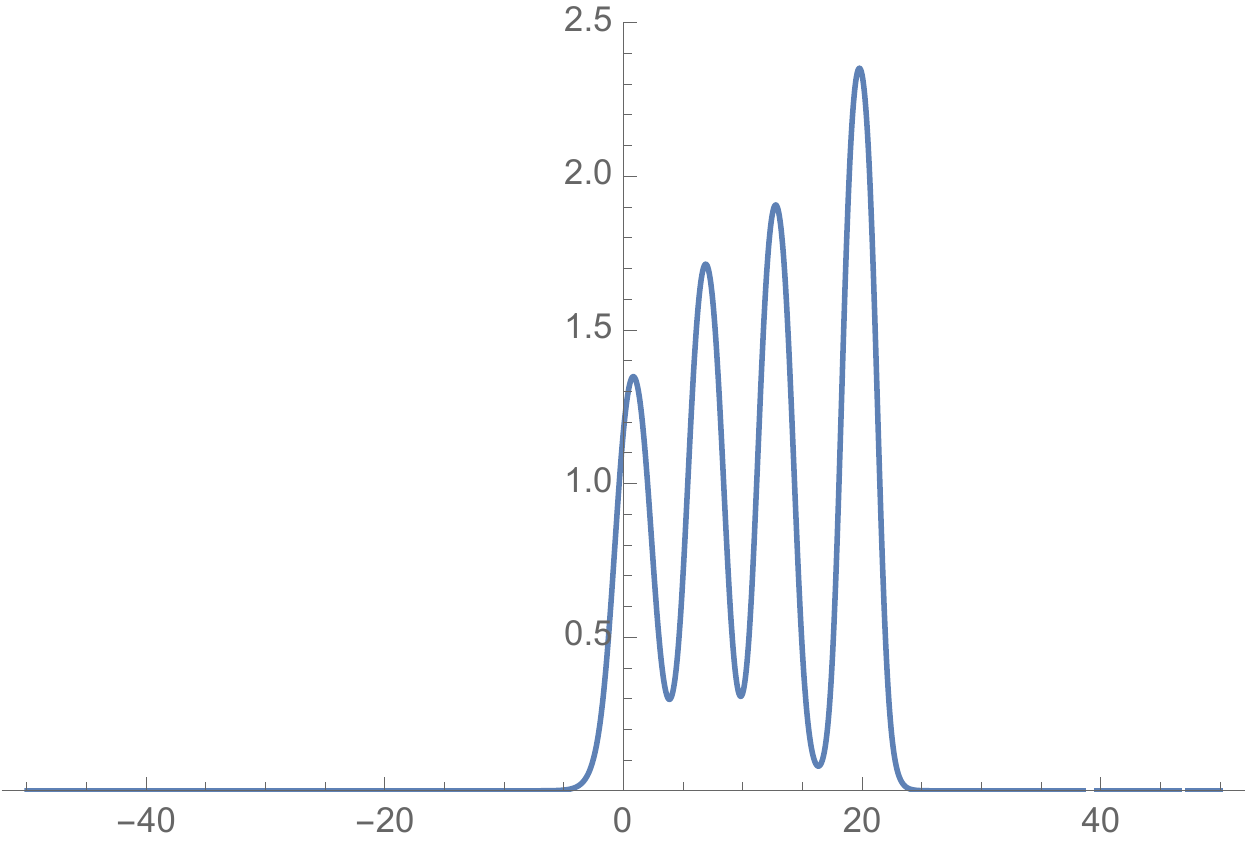}
\caption{$t^{1,1}=-4$}\label{fig-2}
\end{minipage}

\vskip 1em

\begin{minipage}[t]{0.5\linewidth}
\centering
\includegraphics[width=0.8\textwidth]{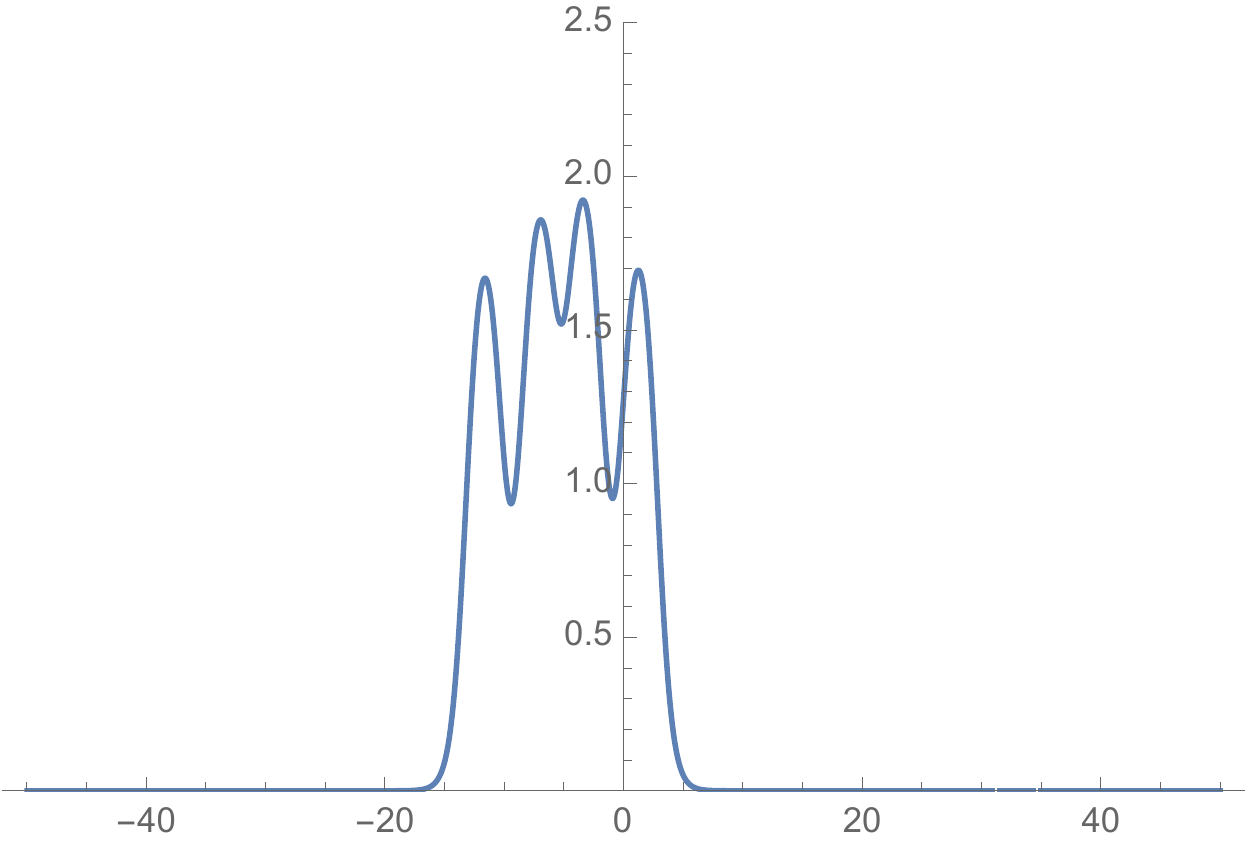}
\caption{$t^{1,1}=0$}\label{fig-3}
\end{minipage}
\begin{minipage}[t]{0.5\linewidth}
\centering
\includegraphics[width=0.8\textwidth]{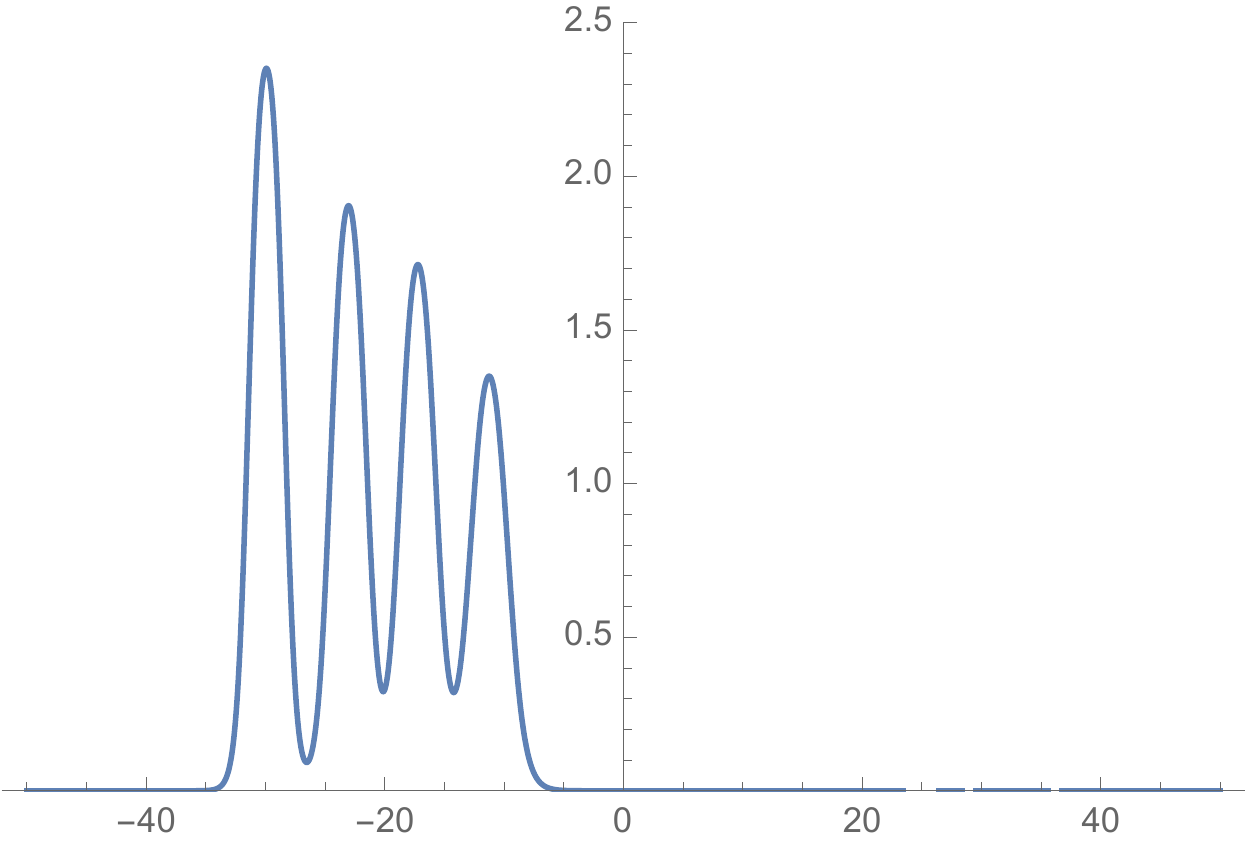}
\caption{$t^{1,1}=4$}\label{fig-4}
\end{minipage}
\end{figure}
\end{emp}

\section{Conclusion}\label{sec-5}

The fractional Volterra hierarchy that we constructed in this paper is a generalization of the usual Volterra lattice hierarchy.
We can also view this hierarchy as a certain reduction of the 2D Toda lattice hierarchy \cite{UT}.
Indeed, the operators $A_i\ (i=1, 2)$ introduced in Lemma \ref{lem-trivial} satisfy the following equations:
\[\e \frac{\p A_i}{\p t^{1,k}}= \left[\left(A_1^k\right)_+, A_i\right],\quad \e \frac{\p A_i}{\p t^{2,k}}=-\left[\left(A_2^k\right)_-, A_i\right],\]
so the FVH is actually a reduction of the 2D Toda lattice hierarchy with the shift operator $\lm_3=e^{\Delta \e \p_x}$. The reduction condition is given by
\[A_1^\beta=A_2^\alpha,\quad \alpha+\beta=\Delta,\]
where $\alpha$, $\beta$ are two nonzero complex numbers. Furthermore, one can consider the following
more general reduction condition:
\[A_1^\beta=A_2^\alpha,\quad \alpha+\beta=m \Delta,\]
where $m$ is a positive integer. This condition will lead to a fractional generalization of the bigraded Toda lattice hierarchy \cite{Car}.

In order to prove Conjecture \ref{main-conj}, we need to study properties of the Virasoro symmetries of the FVH.
We will do it in a subsequent paper.

\vskip 1em
\noindent \textbf{Acknowledgement} We are grateful to Boris Dubrovin and Di Yang for sharing with us
their discovery of the relation of special cubic Hodge integrals with the equation \eqref{new-hie-1} and for helpful discussions. We would also like to thank
Fedor Petrov and Vladimir Dotsenko for their proof of the four identities given in the end of Sect.\,\ref{sec-3}.
This work is supported by NSFC No. 11371214 and No.11471182.

\end{document}